\documentclass{article}

\usepackage{arxiv}

\usepackage[utf8]{inputenc}
\usepackage[T1]{fontenc}
\usepackage{hyperref}
\usepackage{url}
\usepackage{booktabs}
\usepackage{amsfonts}
\usepackage{nicefrac}
\usepackage{microtype}
\usepackage{graphicx}
\usepackage{caption}
\usepackage{subcaption}
\usepackage{tikz-cd}
\usetikzlibrary{arrows}
\usepackage{amsmath}
\usepackage{comment}
\usepackage{amsthm}
\usepackage{amssymb}
\usepackage{url}
\usepackage{tikz}
\usepackage{tikz-qtree,tikz-qtree-compat}
\usetikzlibrary{positioning}
\usepackage[ruled,vlined]{algorithm2e}

\newtheorem{theorem}{Theorem}
\newtheorem{definition}[theorem]{Definition}

\newcommand\independent{\protect\mathpalette{\protect\independenT}{\perp}}
\def\independenT#1#2{\mathrel{\rlap{$#1#2$}\mkern2mu{#1#2}}}

\usepackage{stackengine}
\def\delequal{\mathrel{\ensurestackMath{\stackon[1pt]{=}{\scriptstyle\Delta}}}}

\title{Causes of Effects: Learning individual responses from population data}

\author{
  Scott Mueller\\
  University of California, Los Angeles\\
  Computer Science Department\\
  \texttt{scott@cs.ucla.edu} \\
   \And
 Ang Li\\
  University of California, Los Angeles\\
  Computer Science Department\\
  \texttt{angli@cs.ucla.edu} \\
   \AND
 Judea Pearl\\
  University of California, Los Angeles\\
  Computer Science Department\\
  \texttt{judea@cs.ucla.edu} \\
}

\begin{document}
\maketitle

\begin{abstract}
The problem of individualization is recognized as crucial in almost every field. Identifying causes of effects in specific events is likewise essential for accurate decision making. However, such estimates invoke counterfactual relationships, and are therefore indeterminable from population data. For example, the probability of benefiting from a treatment concerns an individual having a favorable outcome if treated \emph{and} an unfavorable outcome if untreated. Experiments conditioning on fine-grained features are fundamentally inadequate because we can't test both possibilities for an individual. Tian and Pearl provided bounds on this and other probabilities of causation using a combination of experimental and observational data. Even though those bounds were proven tight, narrower bounds, sometimes significantly so, can be achieved when structural information is available in the form of a causal model. This has the power to solve central problems, such as explainable AI, legal responsibility, and personalized medicine, all of which demand counterfactual logic. We analyze and expand on existing research by applying bounds to the probability of necessity and sufficiency (PNS) along with graphical criteria and practical applications.
\end{abstract}

\section{INTRODUCTION}
\label{introduction}
Machine learning advances have enabled tremendous capabilities of learning functions accurately and efficiently from enormous quantities of data. These functions allow for better policies, like whether a surgery, chemotherapy, or radiation therapy is most effective for a population of given characteristics such as age, sex, and type of symptoms. However, this mapping from characteristics to efficacy can be quite misleading when applied to individual decision making, even when the data originate from a randomized controlled trial (RCT). To see why let’s follow the example treated in \cite{mueller2020}. Imagine a novel vaccine for a deadly virus in the midst of a pandemic is in short supply. We want to administer the vaccine to people most likely to benefit from it. In other words, we need to identify the group most likely to \emph{both} survive if vaccinated and succumb if unvaccinated.

A clinical study is conducted to test the effectiveness of the vaccine. A machine learning algorithm trained on data from this RCT learns a correlation between age and recovery. For simplicity, let's assume a binary age classification: sixty years old and under and over sixty years old. Older people survive $57\%$ of the time when vaccinated and $37\%$ of the time when unvaccinated, while younger people survive $55\%$ of the time when vaccinated and $45\%$ of the time when unvaccinated. A naive interpretation is that the vaccine is $20 - 10 = 10$ percentage points more effective for older people.

Before deciding to vaccinate the elders, we need to assess the percentage of elderly patients who would actually benefit from the treatment (PNS) and compare it to the percentage of beneficiaries among the young. Such assessment requires counterfactual analysis such as in \cite{tian2000probabilities} and, based on the data above, yields the following bounds: the probability of over-sixties benefiting from the vaccine is between $20\%$ and $57\%$, while the under-sixties' probability is between $10\%$ and $55\%$. We see that it's anything but clear which group should be vaccinated first.

What is more remarkable is these bounds can be narrowed significantly if data from observational studies is also available, and may even flip priority from the elderly to the young. Observational studies reflect individuals' willingness to get vaccinated in the two age groups. In our example, one can show that the
the bounds for over-sixties and under-sixties may become $[20\%, 40\%]$ and $[40\%, 55\%]$ respectively, thus reversing the na\"{i}ve priorities above.

Clearly, when a subpopulation with a particular set of characteristics is analyzed for PNS, those covariates should be conditioned on. However, this is not always possible, as in the case of ancestral knowledge or a mediating effect of a vaccine. We may have data on the population, but not at the level an individual can make a decision from. After all, the individual doesn't know a potential side-effect of the vaccine until after it's been administered. We present a method to potentially obtain narrower bounds by utilizing population-level data and mild structural assumptions. 

Since Tian and Pearl \cite{tian2000probabilities}, the problem of bounding probabilities of causation was analyzed by combining only two sources of information: experimental data and observational studies. It's surprising that knowing the structure of the causal graph allows us to narrow the bounds, despite the fact that the graph may seem redundant; i.e., we already know the causal effects.  Moreover, the graph adds information about an individual, although it describes properties of the population. The analysis of causes of effects can now take advantage of the causal diagram.

\section{PRELIMINARIES AND RELATED
WORK}
\label{related work}
In this section, we review the definitions for the three aspects of causation as defined in \cite{pearl1999probabilities}. We use the causal diagrams \cite{pearl1995causal, spirtes2000causation, pearl2009causality, koller2009probabilistic} and the language of counterfactuals in its structural model semantics, as given in \cite{balke2013counterfactuals, galles1998axiomatic, halpern2000axiomatizing}. 

We use $Y_x=y$ to denote the counterfactual sentence ``Variable $Y$ would have the value $y$, had $X$ been $x$". For simplicity purposes, in the rest of the paper, we use $y_x$ to denote the event $Y_x=y$, $y_{x'}$ to denote the event $Y_{x'}=y$, $y'_x$ to denote the event $Y_x=y'$, and $y'_{x'}$ to denote the event $Y_{x'}=y'$. For notational simplicity, we limit the discussion to binary $X$ and $Y$, extension to multi-valued variables are straightforward \cite{pearl2009causality}.



Three prominent probabilities of causation are the following:
\vspace{10pt}
\begin{definition}[Probability of necessity (PN)]
Let $X$ and $Y$ be two binary variables in a causal model $M$, let $x$ and $y$ stand for the propositions $X=true$ and $Y=true$, respectively, and $x'$ and $y'$ for their complements. The probability of necessity is defined as the expression \cite{pearl1999probabilities}\\
\begin{eqnarray}
\text{PN} & \delequal & P(Y_{x'}=false|X=true,Y=true)\nonumber \\ 
& \delequal & P(y'_{x'}|x,y)
\label{pn}
\end{eqnarray}
\end{definition}
\par
In other words, PN stands for the probability that event $y$ would not have occurred in the absence of event $x$, given that $x$ and $y$ did in fact occur.

Note that lower case letters (e.g., $x,y$) stand for propositions (or events). PN has applications in epidemiology, legal reasoning, and artificial intelligence. Epidemiologists have long been concerned with estimating the probability that a certain case of disease is attributable to a particular exposure, which is normally interpreted counterfactually as ``the probability that disease would not have occurred in the absence of exposure, given that disease and exposure did in fact occur." This counterfactual notion is also used frequently in lawsuits, where legal responsibility is at the center of contention.
\vspace{10pt}
\begin{definition}[Probability of sufficiency (PS)] \cite{pearl1999probabilities}
\begin{eqnarray}
\text{PS}\delequal P(y_x|y',x')
\label{ps}
\end{eqnarray}
\end{definition}

PS finds applications in policy analysis, artificial intelligence, and psychology.
A policy maker may well be interested in the dangers that a certain exposure may present to the healthy population \cite{khoury1989measurement}. Counterfactually, this notion is expressed as the ``probability that a healthy unexposed individual would have gotten the disease had he/she been exposed." In psychology, PS serves as the basis for Cheng's \cite{cheng1997covariation} causal power theory \cite{glymour2013psychological}, which attempts to explain how humans judge causal strength among events. In artificial intelligence, PS plays a major role in the generation of explanations \cite{pearl2009causality}.
\vspace{10pt}
\begin{definition}[Probability of necessity and sufficiency (PNS)] \cite{pearl1999probabilities}
\begin{eqnarray}
\text{PNS}\delequal P(y_x,y'_{x'})
\label{pns}
\end{eqnarray}
\end{definition}
\par
PNS stands for the probability that $y$ would respond to $x$ both ways, and therefore measures both the sufficiency and necessity of $x$ to produce $y$.

Tian and Pearl \cite{tian2000probabilities} provide tight bounds for PNS, PN, and PS without a causal diagram using Balke's program \cite{balke1997probabilistic}. Li and Pearl \cite{li2019unit} provide a theoretical proof of the tight bounds for PNS, PS, PN, and other probabilities of causation without a causal diagram.

PNS, PN, and PS have the following tight bounds:\\

\begin{eqnarray}
\max \left \{
\begin{array}{cc}
0 \\
P(y_x) - P(y_{x'}) \\
P(y) - P(y_{x'}) \\
P(y_x) - P(y)\\
\end{array}
\right \}
\le \text{PNS}
\label{pnslb}
\end{eqnarray}

\begin{eqnarray}
\text{PNS} \le \min \left \{
\begin{array}{cc}
 P(y_x) \\
 P(y'_{x'}) \\
P(x,y) + P(x',y') \\
P(y_x) - P(y_{x'}) +\\
+ P(x, y') + P(x', y)
\end{array} 
\right \}
\label{pnsub}
\end{eqnarray}

\begin{eqnarray}
\max \left \{
\begin{array}{cc}
0 \\
\frac{P(y)-P(y_{x'})}{P(x,y)}
\end{array} 
\right \}
\le \text{PN}
\label{pnlb}
\end{eqnarray}

\begin{eqnarray}
\text{PN} \le
\min \left \{
\begin{array}{cc}
1 \\
\frac{P(y'_{x'})-P(x',y')}{P(x,y)} 
\end{array}
\right \}
\label{pnub}
\end{eqnarray}

Note that we only consider PNS and PN here because the bounds of PS can be easily obtained by exchanging $x$ with $x'$ and $y$ with $y'$ in the bounds of PN.

To obtain bounds for a specific population, defined by a set $C$ of characteristics, the expressions above should be modified by conditioning each term on $C=c$. This would normally yield narrower bounds because, when $C$ is not affected by $X$, it reduces variations among units in the subpopulation considered. In this paper, however, we obtain narrower bounds of PNS by leveraging another source of knowledge -- the causal diagram behind the data, together with measurements of a set $Z$ of covariates in that diagram. We provide graphical conditions under which the availability of such measurements would improve the bounds and demonstrate, both analytically and by simulation, the degree of improvement achieved. Narrower bounds and graphical criteria can be obtained for PN and PS through the same mechanism detailed in the proofs in the appendix.

\section{BOUNDS WITH CAUSAL DIAGRAM}
\label{bounds}
\subsection{With additional covariate $Z$}
\subsubsection{Non-descendant $Z$}
Theorems \ref{thm1} and \ref{thm2} below provide bounds for PNS when a set $Z$ of variables can be measured which satisfy only one simple condition: $Z$ contains no descendant of $X$. This condition is important  because if $X$ was set to $x$ and $Z$ contains a descendant of $X$, then $Z$ could be altered as well and $P(y_x|z)$ would be unmeasurable. If the descendant is independent of $Y_x$, then $P(y_x|z$ would be measurable, but that descendant wouldn't contribute to any narrowing of bounds. These bounds are always contained within the Tian-Pearl bounds of equations \ref{pnslb}, \ref{pnsub}, \ref{pnlb}, and \ref{pnub}.

\begin{theorem}
Given a causal diagram $G$ and distribution compatible with $G$, let $Z$ be a set of variables that does not contain any descendant of $X$ in $G$, then PNS is bounded as follows:
\begin{flushleft}
\begin{eqnarray}
\sum_z\max\left\{
\begin{array}{c}
0,\\
P(y_x|z)-P(y_{x'}|z),\\
P(y|z)-P(y_{x'}|z),\\
P(y_x|z)-P(y|z)
\end{array}
\right\}\times P(z)\le \text{PNS}
\label{inequ11}
\end{eqnarray}
\end{flushleft}
\begin{flushleft}
\begin{equation}
\sum_z\min\left\{
\begin{array}{c}
P(y_x|z),\\
P(y'_{x'}|z),\\
P(y,x|z)+P(y',x'|z),\\
P(y_x|z)-P(y_{x'}|z)+\\
+P(y,x'|z)+P(y',x|z)
\end{array}
\right\}\times P(z)\ge \text{PNS}
\label{inequ22}
\end{equation}
\end{flushleft}
\label{thm1}
\begin{proof}
See Appendix.
\end{proof}
\end{theorem}

Note that, unlike the population-specific bounds, where each term was conditioned on $C=c$, here $Z=z$ enters only some of the terms. This is because the measurement of $Z$ is conducted in the study, but may not be available for the individual seeking advice. Examples are illustrated in Section \ref{examples}.

Note also that if only experimental data are available (i.e., $P(Y), P(Y, X), P(Y|Z), P(Y, X|Z)$ are not measured), arguments to the max or min functions involving observational data can be disregarded. For example, the lower bounds of theorem \ref{thm1} would become $\max \{P(y_x)-P(y_{x'}), \sum_z \max\{0,P(y_x|z)-P(y_{x'}|z)\}\times P(z)\}$.

\subsubsection{Sufficient Covariate $Z$}

\begin{figure}[ht]
\centering
\begin{subfigure}{0.49\textwidth}
\centering
\begin{tikzpicture}[->,>=stealth',node distance=1cm,
  thick,main node/.style={circle,fill,inner sep=1.5pt}]
  \node[main node] (1) [label=above:{$Z$}]{};
  \node[main node] (2) [below left =1cm of 1,label=left:$X$]{};
  \node[main node] (3) [below right =1cm of 1,label=right:$Y$] {};
  \path[every node/.style={font=\sffamily\small}]
    (1) edge node {} (2)
    (1) edge node {} (3)
    (2) edge node {} (3);
\end{tikzpicture}
\caption{Confounder $Z$}
\label{causalg2}
\end{subfigure}
\hfill
\begin{subfigure}{0.49\textwidth}
\centering
\begin{tikzpicture}[->,>=stealth',node distance=1cm,
  thick,main node/.style={circle,fill,inner sep=1.5pt}]
  \node[main node] (1) [label=above:{$Z$}]{};
  \node[main node] (2) [below left =1cm of 1,label=left:$X$]{};
  \node[main node] (3) [below right =1cm of 1,label=right:$Y$] {};
  \path[every node/.style={font=\sffamily\small}]
    (1) edge node {} (3)
    (2) edge node {} (3);
\end{tikzpicture}
\caption{Outcome-affecting covariate $Z$}
\label{causalg3}
\end{subfigure}
\caption{$Z$ is not a descendant of $X$}
\label{causalcovariate}
\end{figure}
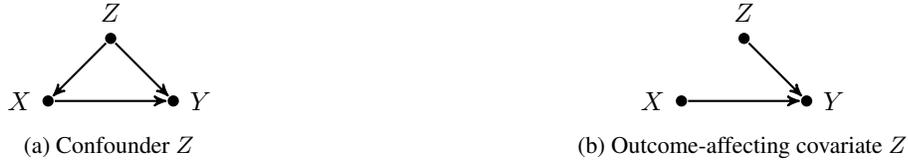

In figures \ref{causalg2} and \ref{causalg3}, $Z$ is not a descendant of $X$ and further satisfies the back-door criterion. For such cases the PNS bounds can be simplified to read:
\vspace{10pt}
\begin{theorem}
Given a causal diagram $G$ and distribution compatible with $G$, let $Z$ be a set of variables satisfying the back-door criterion \cite{pearl1993aspects} in $G$, then the PNS is bounded as follows:
\begin{flushleft}
\begin{eqnarray}
\sum_z \max\{0,P(y|x,z)-P(y|x',z)\}\times P(z) \le \text{PNS}
\label{lb_confounder}
\end{eqnarray}
\end{flushleft}
\begin{flushleft}
\begin{eqnarray}
\sum_z \min\{P(y|x,z),P(y'|x',z)\}\times P(z)
\ge \text{PNS}
\label{ub_confounder}
\end{eqnarray}
\end{flushleft}
\begin{proof}
See Appendix.
\end{proof}
\label{thm2}
\end{theorem}
The significance of theorem \ref{thm2} is due to the ability to compute bounds using purely observational data.
\subsection{Mediation}
\subsubsection{Z as a PARTIAL MEDIATOR}
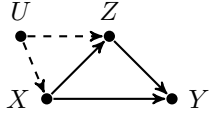
\begin{figure}[h]
\centering
\begin{tikzpicture}[->,>=stealth',node distance=2cm,
  thick,main node/.style={circle,fill,inner sep=1.5pt}]
  \node[main node] (0) [label=above:{$U$}]{};
  \node[main node] (1) [right =1cm of 0,label=above:{$Z$}]{};
  \node[main node] (2) [below left =1cm of 1,label=left:$X$]{};
  \node[main node] (3) [below right =1cm of 1,label=right:$Y$] {};
  \path[every node/.style={font=\sffamily\small}]
    (2) edge node {} (1)
    (1) edge node {} (3)
    (2) edge node {} (3);
  \draw [dashed] (0) -- (1);
  \draw [dashed] (0) -- (2);
\end{tikzpicture}
\caption{Mediator $Z$ with direct effect}
\label{causalg1}
\end{figure}
In figure \ref{causalg1}, $Z$ is a descendant of $X$, so we cannot use theorems \ref{thm1} and \ref{thm2}. However, the absence of confounders between $Z$ and $Y$ and between $X$ and $Y$ permits us to bound PNS as follows:
\vspace{10pt}
\begin{theorem}
Given a causal diagram $G$ and distribution compatible with $G$, let $Z$ be a set of variables such that $\forall x,x' \in X : x \ne x', (Y_x \independent X\cup Z_{x'}\ |\ Z_x)$ in $G$, then the PNS is bounded as follows:
\begin{flushleft}
\begin{eqnarray}
\max\left\{
\begin{array}{c}
0,\\
P(y_x)-P(y_{x'}),\\
P(y)-P(y_{x'}),\\
P(y_x)-P(y)\\
\end{array}
\right\}\le \text{PNS}
\label{lb_mediator_plus_direct}
\end{eqnarray}
\end{flushleft}

\begin{flushleft}
\begin{eqnarray}
\min\left\{
\begin{array}{c}
P(y_x),\\
P(y'_{x'}),\\
P(y,x)+P(y',x'),\\
\\
P(y_x)-P(y_{x'})+\\
+P(y,x')+P(y',x),\\
\\
\sum_z \sum_{z'} \min\{P(y|z,x),\\
P(y'|z',x')\}\times\\
\min\{P(z_x),P(z'_{x'})\}
\end{array}
\right\}\ge \text{PNS}
\label{ub_mediator_plus_direct}
\end{eqnarray}
\end{flushleft}
\label{thm3}
\begin{proof}
See Appendix.
\end{proof}
\end{theorem}

Note that although this lower bound is unchanged from Tian and Pearl, the upper bound contains a vital additional argument to the min function. This new term can significantly reduce the upper bound. The rest of the terms are included because sometimes Tian and Pearl's bounds are superior. The following theorem has the same quality.
\subsubsection{PURE MEDIATOR}
Figure \ref{causalg5} is a special case of figure \ref{causalg1}, in which $X$ has no direct effect on $Y$. The resulting bounds for PNS read:
\vspace{10pt}
\begin{theorem}
Given a causal diagram $G$ in figure \ref{causalg5} and distribution that compatible with $G$, then PNS are bounded as follow:

\begin{figure}[h]
\centering
\begin{tikzpicture}[->,>=stealth',node distance=2cm,
  thick,main node/.style={circle,fill,inner sep=1.5pt}]
  \node[main node] (1) [label=above:{$Z$}]{};
  \node[main node] (2) [below left =1cm of 1,label=left:$X$]{};
  \node[main node] (3) [below right =1cm of 1,label=right:$Y$] {};
  \path[every node/.style={font=\sffamily\small}]
    (2) edge node {} (1)
    (1) edge node {} (3);
\end{tikzpicture}
\caption{Mediator $Z$ with no direct effect}
\label{causalg5}
\end{figure}
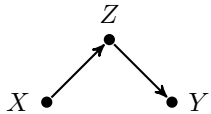

\begin{flushleft}
\begin{eqnarray}
\max\left\{
\begin{array}{c}
0,\\
P(y_x)-P(y_{x'}),\\
P(y)-P(y_{x'}),\\
P(y_x)-P(y)\\
\end{array}
\right\}\le \text{PNS}
\label{lb_mediator}
\end{eqnarray}
\end{flushleft}

\begin{flushleft}
\begin{eqnarray}
\min\left\{
\begin{array}{c}
P(y_x),\\
P(y'_{x'}),\\
P(y,x)+P(y',x'),\\
\\
P(y_x)-P(y_{x'})+\\
+P(y,x')+P(y',x),\\
\\
\Sigma_z \Sigma_{z'\ne z}\min\{P(y|z),P(y'|z')\}\times\\
\min\{P(z|x),P(z'|x')\}
\end{array}
\right\}\ge \text{PNS}
\label{ub_mediator}
\end{eqnarray}
\end{flushleft}
\label{thm4}
\begin{proof}
See Appendix.
\end{proof}
\end{theorem}

The core terms for theorems \ref{thm3} and \ref{thm4} added to the upper bounds notably only require observational data.

\section{EXAMPLES}
\label{examples}

\subsection{CREDIT TO THE TREATMENT}
The manufacturer of a drug wants to claim that a non-trivial number of recovered patients who were given access to the drug owe their recovery to the drug. So they conduct an observational study; they record the recovery rates of $700$ patients. $464$ patients chose to take the drug and $236$ patients did not. The results of the study are in table \ref{obsdata}. The manufacturer claims success for their drug because the overall recovery rate from the observational study has increased from $54\%$ to $68\%$ for non-drug-takers to drug-takers.

\begin{table}[h]
\centering
\begin{tabular}{|c|c|c|}
\hline 
&Drug&No Drug\\
\hline
Women&\begin{tabular}{c}$1$ out of $110$ \\recovered ($1\%$)\end{tabular}&\begin{tabular}{c}$13$ out of $120$ \\recovered ($11\%$)\end{tabular}\\
\hline
Men&\begin{tabular}{c}$313$ out of $354$ \\recovered ($88\%$)\end{tabular}&\begin{tabular}{c}$114$ out of $116$ \\recovered ($98\%$)\end{tabular}\\
\hline
Overall&\begin{tabular}{c}$314$ out of $464$ \\recovered ($68\%$)\end{tabular}&\begin{tabular}{c}$127$ out of $236$ \\recovered ($54\%$)\end{tabular}\\
\hline
\end{tabular}
\vspace{5pt}
\caption{Results of a drug study with gender taken into account}
\label{obsdata}
\end{table}

The number of recovered patients that should credit the drug for their recovery are those who would recover if they had taken the drug and would not recover if they had not taken the drug. This is the PNS.

Let $X=x$ denote the event that the patient took the drug and $X=x'$ denote the event that the patient did not take the drug. Let $Y=y$ denote the event that the patient has recovered and $Y=y'$ denote the event that the patient has not recovered. Let $Z=z$ represent female patients and $Z=z'$ represent male patients. Suppose we know an additional fact, estrogen has a negative effect on recovery, so women are less likely to recover than men, regardless of the drug. Additionally, as we can see from the data, men are significantly more likely to take the drug than women are. The causal diagram is shown in Figure \ref{causalg2}.

Node $Z$ on the graph satisfies the back-door criterion, therefore we can compute the causal effect $P(y_x)$ and $P(y_x')$ via the adjustment formula \cite{pearl1993aspects} and observational data from table \ref{obsdata}, where,
\begin{eqnarray*}
P(y_x) = \sum\limits_z P(y|x,z)P(z)=0.597,\\
P(y_{x'}) = \sum\limits_z P(y|x',z)P(z)=0.696,\\
P(y'_{x'}) = 1 - P(y_{x'}) = 0.304.
\end{eqnarray*}
\par
Therefore, the bounds of PNS computed using equations \ref{pnslb} and \ref{pnsub} are $0\le PNS\le 0.297$, where the diagram was used only to identify the causal effects $y_x$ and $y_{x'}$. These bounds aren't informative enough to conclude whether or not the drug was the cause of recovery for a meaningful number of patients. They suggest that the fraction of beneficiaries can be as low as $0\%$ or as high as $29.7\%$. Now, consider the bounds in theorem \ref{thm2} which takes into account the position of $Z$ in the diagram.
Since $Z$ satisfies the back-door criterion, we can use equations \ref{lb_confounder} and \ref{ub_confounder} to compute $0 \leqslant PNS \leqslant 0.01$. The conclusion now is obvious. At most $7$ out of $314$ patients' recoveries can be credited to the drug. This is strong evidence that counters the manufacturer's claim.

\subsection{INFLAMMATION MEDIATOR}
As before, let $X$ and $Y$ represent drug consumption and recovery. Let $Z$ represent acute inflammation with $z$ being present and $z'$ being absent. In some people, the drug causes acute inflammation, which has adverse effects on recovery, so the causal structure is depicted in figure \ref{causalg5}. We observe the following proportions among drug takers, non-takers, with inflammation, and without inflammation:
\begin{align*}
    \begin{split}
    P(y|z) &= 0.5,\\
    P(y|z') &= 0.5,
    \end{split}
    \begin{split}
    P(z|x) &= 0.1,\\
    P(z|x') &= 0.1.
    \end{split}
\end{align*}
\par
The Tian-Pearl PNS upper bound is:
\begin{align*}
    \text{PNS} \leqslant \min{\{P(y|x), P(y'|x')\}} = 0.5\text.
\end{align*}
\par
Given that the lower bound is $0$, these bounds are not very informative. If we knew that an individual would react to the drug with acute inflammation, we would only look at the data comprising of people reacting to the drug with acute inflammation. Since we are conditioning on $z$, $PNS = 0$ because the outcome, $Y$, will have the same result regardless of whether the person consumed the drug. So knowing a person's inflammation response to the drug narrows PNS from a wide $[0, 0.5]$ to a point estimate of $0$. Imagine, for this drug, that we can't know ahead of time how a person will react inflammation-wise. We can only observe acute inflammation after the drug is administered. Since we have population data from patients who have already taken the drug, we can utilize this mediator to bound the PNS for new patients who haven't yet taken the drug:
\begin{align*}
    \text{PNS} &\leqslant \min{\left\{\begin{array}{lr}
P(y|z) \cdot P(z|x) + P(y|z') \cdot P(z|x'),\\
P(y|z) \cdot P(z'|x') + P(y|z') \cdot P(z'|x),\\
P(y'|z') \cdot P(z|x) + P(y'|z) \cdot P(z|x'),\\
P(y'|z') \cdot P(z'|x') + P(y'|z) \cdot P(z'|x)
\end{array}\right\}}\\
    &= 0.1\text.
\end{align*}
\par
The mediator-improved PNS upper bound is significantly smaller than what the Tian-Pearl upper bound provides, $0.1$ vs $0.5$. The new upper bound can now be effectively weighed against other factors like cost and side-effects.

\subsection{ANCESTRAL COVARIATE}
\label{ancestral_covariate}
Let's continue from the introduction, where $X$ represents vaccination with $x$ being vaccinated and $x'$ being unvaccinated and $Y$ represents survival with $y$ is surviving and $y'$ is succumbing to the pandemic. Instead of classifying by age, let's assume our machine learning algorithm uncovers a correlation between survival and ancestry. Let $Z$ represent ancestry and, for simplicity, there are only two ancestries, $z$ and $z'$. Either graph of figure \ref{causalcovariate} is representative of this. Our RCT data reveals:
\begin{align*}
    \begin{split}
    P(Z = z) &= 0.5\text,\\
    P(y_x|Z = z) &= 0.75\text,\\
    P(y_{x'}|Z = z) &= 0.2\text,
    \end{split}
    \begin{split}
    P(y_x|Z = z') &= 0.25\text,\\
    P(y_{x'}|Z = z') &= 0.6\text.
    \end{split}
\end{align*}
\par
We now have four different bounds on PNS:
\begin{align*}
\text{Tian-Pearl} &\implies 0.1 \leqslant PNS \leqslant 0.5\\
\text{Covariate-improved} &\implies 0.275 \leqslant PNS \leqslant 0.5\\
\text{Person has ancestry }z &\implies 0.55 \leqslant PNS \leqslant 0.75\\
\text{Person has ancestry }z' &\implies 0 \leqslant PNS \leqslant 0.25\\
\end{align*}
As expected, using the causal diagram and ancestral $Z$ yields narrower bounds than the Tian-Pearl bounds. However, it’s surprising that knowing a person has either ancestry $z$ or $z'$ gives us bounds outside of our new bounds. In fact, they are completely outside the wider Tian-Pearl bounds. This is discussed in section \ref{discussion}.

In the meantime, it's important to recognize that the last two ancestry-specific PNS bounds are what should be referred to if an individual knows their ancestry. The covariate-improved PNS bounds should only be referred to if a person's ancestry is unknown. This might be because the person was adopted with no hint as to whether they're from ancestry $z$ or $z'$ (physical features are right in between or indistinguishable).

\section{SIMULATION RESULTS}
\label{simulation}

In this section, we illustrate that the bounds of PNS are improved in four different simple causal diagrams.

The first causal diagram is the simplest one in figure \ref{causalg2} with binary $Z$ satisfying the back-door criterion. The second causal diagram is figure \ref{causalgexample3} where $\{Z_1,Z_2\}$ satisfies the back-door criterion and both $Z_1$ and $Z_2$ are binary. The third causal diagram is figure \ref{causalgexample4}, with six subsets of $Z_1$, $Z_2$, and $Z_3$ satisfying the back-door criterion. The last causal diagram is figure \ref{causalg2}, the same as the first causal diagram, except $Z$ has $1024$ instantiates.

\begin{figure}
\centering
\includegraphics[width=0.45\textwidth]{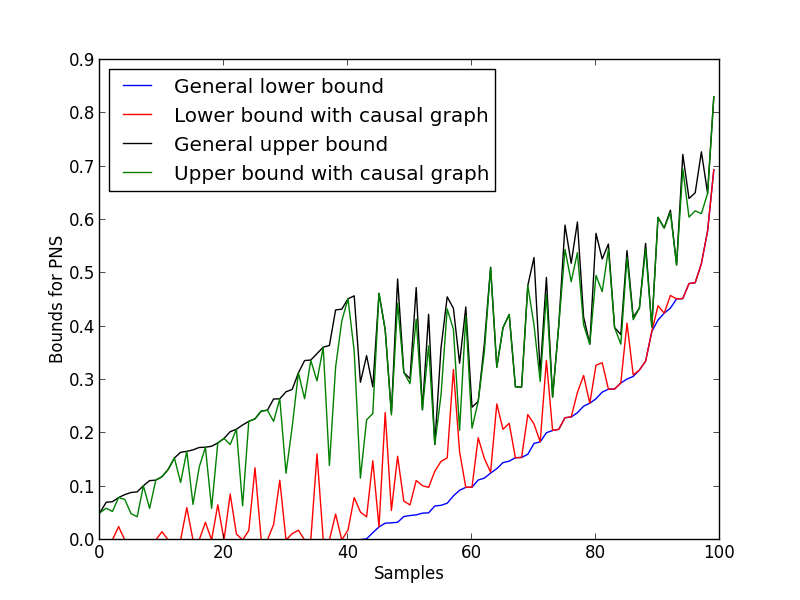}
\caption{Bounds of PNS in Figure \ref{causalg2}, where general bounds are obtained from Equation \ref{pnslb} and \ref{pnsub} and the bounds with causal diagram are obtained by Theorem \ref{thm2}.}
\label{res1}
\end{figure}

\begin{figure}
\centering
\includegraphics[width=0.45\textwidth]{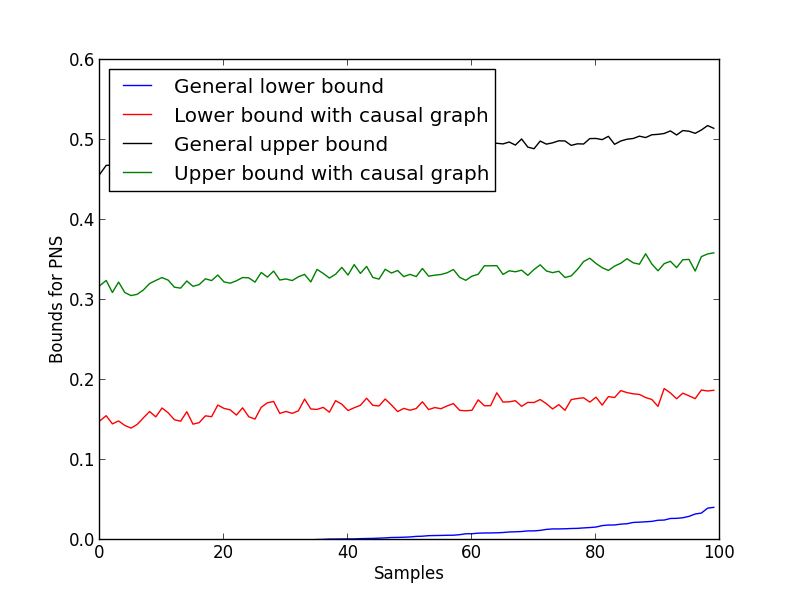}
\caption{Bounds of PNS in figure \ref{causalg2} with $Z$ having $1024$ instantiates, where general bounds are obtained from Equation \ref{pnslb} and \ref{pnsub} and the bounds with causal diagram are obtained by theorem \ref{thm2}.}
\label{res4}
\end{figure}

For each of the causal diagrams, we randomly generated $100000$ sample distributions compatible with the causal diagram. We compared the average increased lower bound (i.e., lower bound with causal diagram - lower bound without causal diagram), the average decreased upper bound (i.e., upper bound without causal diagram - upper bound with causal diagram), the average gap without causal diagram (i.e., upper bound without causal diagram - lower bound without causal diagram), and the average gap with causal diagram (i.e., upper bound with causal diagram - lower bound with causal diagram). The results are summarized in table \ref{simres}. For each of the causal diagrams, we then randomly pick $100$ out of $100000$ sample distributions to draw the graph of bounds with and without causal diagram. The results are in figures \ref{res1} to \ref{res4}. 

\begin{figure}[h]
\centering
\begin{tikzpicture}[->,>=stealth',node distance=2cm,
  thick,main node/.style={circle,fill,inner sep=1.5pt}]
  \node[main node] (1) [label=above:{$Z_1$}]{};
  \node[main node] (2) [below =1cm of 1] [label=above:{$Z_2$}]{};
  \node[main node] (3) [below left =1cm of 2,label=left:$X$]{};
  \node[main node] (4) [below right =1cm of 2,label=right:$Y$] {};
  \path[every node/.style={font=\sffamily\small}]
    (1) edge node {} (3)
    (1) edge node {} (4)
    (2) edge node {} (3)
    (2) edge node {} (4)
    (3) edge node {} (4);
\end{tikzpicture}
\caption{}
\label{causalgexample3}
\end{figure}

\begin{figure}[h]
\centering
\begin{tikzpicture}[->,>=stealth',node distance=2cm,
  thick,main node/.style={circle,fill,inner sep=1.5pt}]
  \node[main node] (1) [label=above:{$Z_1$}]{};
  \node[main node] (2) [below right =1.414cm of 1] [label=above:{$Z_2$}]{};
  \node[main node] (3) [right =2cm of 1] [label=above:{$Z_3$}]{};
  \node[main node] (4) [below =2cm of 1,label=left:$X$]{};
  \node[main node] (5) [below =2cm of 3,label=right:$Y$] {};
  \path[every node/.style={font=\sffamily\small}]
    (1) edge node {} (2)
    (1) edge node {} (4)
    (3) edge node {} (2)
    (3) edge node {} (5)
    (4) edge node {} (5);
\end{tikzpicture}
\caption{}
\label{causalgexample4}
\end{figure}
\begin{table}[h]
\centering
\begin{tabular}{|c|c|c|c|c|}
\hline
&Fig \ref{causalg2}&Fig \ref{causalgexample3}&Fig \ref{causalgexample4} &\begin{tabular}{c}Fig \ref{causalg2}\\ with non-\\binary $Z$\end{tabular}\\
\hline
\begin{tabular}{c}Average\\increased\\lower\\bound\end{tabular}&0.026&0.050&0.032&0.158\\
\hline
\begin{tabular}{c}Average\\decreased\\upper\\bound\end{tabular}&0.026&0.050&0.032&0.158\\
\hline
\begin{tabular}{c}Average\\gap\\without\\ causal\\diagram\end{tabular}&0.219&0.267&0.231&0.483\\
\hline
\begin{tabular}{c}Average\\gap\\with\\causal\\diagram\end{tabular}&0.166&0.166&0.166&0.166\\
\hline
\end{tabular}
\vspace{5pt}
\caption{Bounds of PNS with and without causal diagram}
\label{simres}
\end{table}
\begin{figure}
\centering
\includegraphics[width=0.45\textwidth]{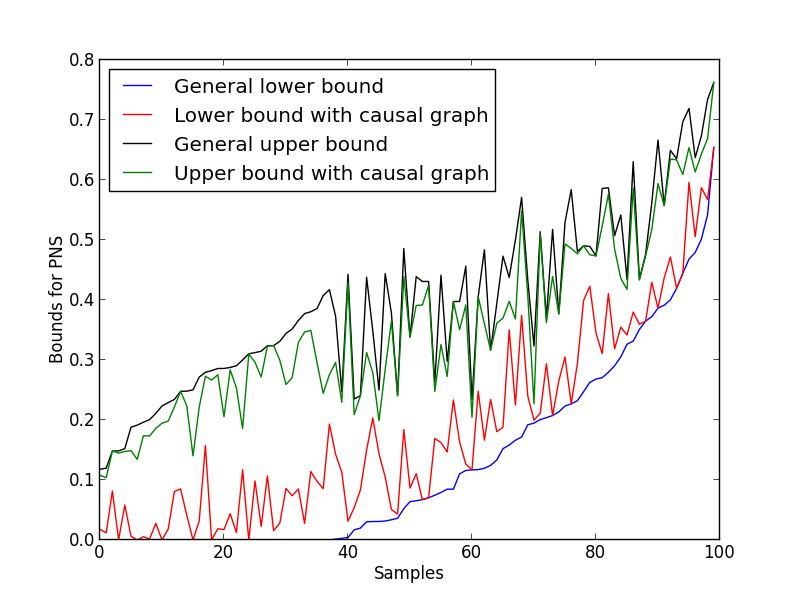}
\caption{Bounds of PNS in figure \ref{causalgexample3}, where general bounds are obtained from equations \ref{pnslb} and \ref{pnsub} and the bounds with causal diagram are obtained by theorem \ref{thm2}.}
\label{res2}
\end{figure}
\begin{figure}
\centering
\includegraphics[width=0.45\textwidth]{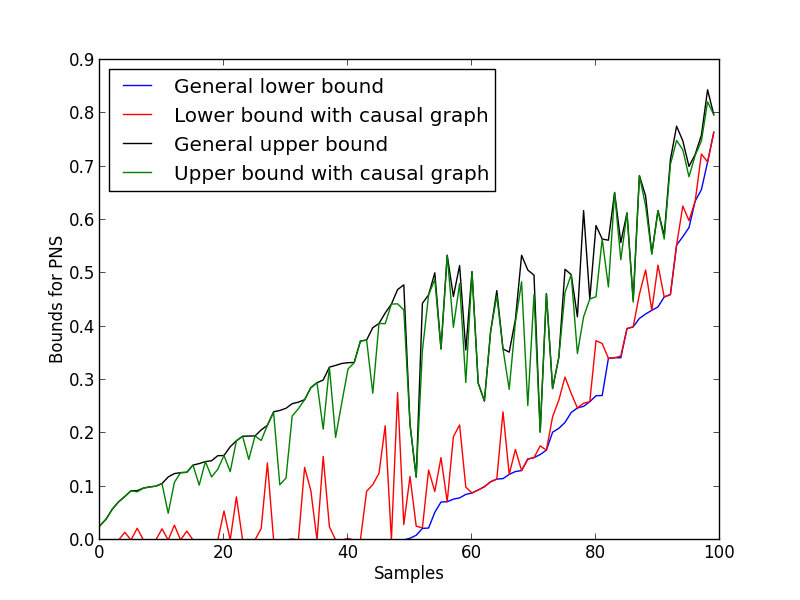}
\caption{Bounds of PNS in figure \ref{causalgexample4}, where general bounds are obtained from equations \ref{pnslb} and \ref{pnsub} and the bounds with causal diagram are obtained by theorem \ref{thm2}.}
\label{res3}
\end{figure}

From figures \ref{res1} to \ref{res4}, we see that the bounds of PNS are improved in most of the samples with each causal diagram. Surprisingly, we see in table \ref{simres} that the average gap without a causal diagram fluctuates substantially (between $0.219$ and $0.483$) between causal diagrams. However, the average gap with a causal diagram is consistently $0.166$ among all causal diagrams. If subsets $Z$ are available that satisfy theorem \ref{thm1}, the bounds of PNS are useful as the gap is narrow and with low variability across causal structures.

\section{DISCUSSION}
\label{discussion}

We have shown that knowledge of a causal structure enables narrower PNS bounds to be estimated, compared with the tight bounds of Tian and Pearl which were derived without such knowledge. However, it must be emphasized that this narrowing is only applicable to individuals when unable to determine their characteristics at decision time. If their $Z$ values are known, the bounds of equations \ref{pnslb} and \ref{pnsub}, conditioned on those values, should be consulted. Example \ref{ancestral_covariate} provides a scenario where people who know their ancestry have very different PNS bounds than people who don't know their ancestry. You would expect the additional information of ancestral knowledge would further narrow the bounds, but they change the bounds to a different non-overlapping range. This violates the heuristic that \emph{additional information} should narrow the bounds or, at worst, not widen them. To rephrase, if you don't know someone's ancestry, the probability they benefit from this drug is between $0.275$ and $0.5$. Once you acquire the additional information that the person is of ancestry $z$, the probability they benefit from this treatment becomes between $0.55$ and $0.75$. How is this possible? Was the person's probability of benefiting never really between $0.275$ and $0.5$ that we calculated before knowing their ancestry?

The reason for this seeming inconsistency is that we're asking different questions. When we didn't know the ancestry, we were asking, ``what is the probability of benefiting for a person regardless of ancestry?'' When we found out the person is of ancestry $z$, we then asked a different question, ``what is the probability of benefiting for a person of ancestry $z$?'' The additional information of the person's ancestry didn't help the first question and the second question isn't answerable without the additional information.

The following example will illuminate the reasons for this phenomenon \cite[p.~296]{pearl2009causality}.  Let the covariate $Z$ stand for the outcome of a fair coin toss, so $P(Z = \text{heads}) = 0.5$. Without knowing what treatment $X$ and success $Y$ represent, let's assume the following measurements are taken:
\begin{align*}
    \begin{split}
    &P(y_x) = 0.5\text,\\
    &P(y_x|Z = \text{heads}) = 1\text,\\
    &P(y_x|Z = \text{tails}) = 0\text,
    \end{split}
    \begin{split}
    &P(y_{x'}) = 0.5\text,\\
    &P(y_{x'}|Z = \text{heads}) = 0\text,\\
    &P(y_{x'}|Z = \text{tails}) = 1\text.
    \end{split}
\end{align*}
\par
Tian-Pearl bounds gives us $0 \leqslant PNS \leqslant 0.5$ and the bounds utilizing $Z$ are $0.5 \leqslant PNS \leqslant 0.5$ or $\text{PNS} = 0.5$.

Now, let's uncover the functional mechanism, $x$ represents betting \$1 on heads, $x'$ represents betting \$1 on tails, $y$ represents winning \$1, and $y'$ represents losing \$1. It should now be clear why $P(y_x) = P(y_{x'}) = 0.5$. Without knowing the coin toss result, $Z$, the odds of winning \$1 are $50/50$ whether you bet on heads or tails. PNS is also $0.5$ because benefiting from betting on heads is true only when the coin toss was heads. The coin toss is heads $50\%$ of the time.

This brings us back to the PNS bounds when we have the additional information of what the coin toss result was. If we know the coin toss resulted in heads, then the probability of benefiting from betting on heads is $100\%$. Similarly, if we know the coin toss resulted in tails, then the probability of benefiting from betting on heads is $0\%$. In other words $\text{PNS}(\text{heads}) = 1$ and $\text{PNS}(\text{tails}) = 0$. If the coin toss is heads, winning only happens when betting on heads. Even though the bounds are completely different when we provided with the very useful additional information of the coin toss, there is clearly no contradiction here. There was a $50\%$ probability of benefiting from betting on heads when we didn't know the coin toss result and a $100\%$ probability of benefiting from betting on heads when we knew the coin toss resulted in heads. We were asking two separate questions. The first question was, ``what is the probability of benefiting regardless of coin toss result?'' The second question was, ``what is the probability of benefiting for a coin toss of heads?''

\section{CONCLUSION}
\label{conclusion}

In this work, we have developed a graphical method of learning individualized functions (representing PNS, PN, and PS) from population data, based the structure of the causal graph. This generalizes both the PN bounds derived in \cite{dawid2017} and those derived in \cite{tian2000probabilities}. Often these functions return bounds as opposed to point estimates. Nevertheless, these bounds can be tremendously informative. Machine learning algorithms need to incorporate these techniques in order to understand, interpret, and apply the underlying probabilities of causation of their data. Identifying causes of effects and decision making benefit greatly from using population data for individual cases.

\bibliographystyle{apalike}
\bibliography{main}

\begin{thebibliography}{}

\bibitem[Balke and Pearl, 1997]{balke1997probabilistic}
Balke, A. and Pearl, J. (1997).
\newblock {\em Probabilistic Counterfactuals: Semantics, Computation, and
  Applications}.
\newblock PhD thesis, University of California, Los Angeles.

\bibitem[Balke and Pearl, 2013]{balke2013counterfactuals}
Balke, A. and Pearl, J. (2013).
\newblock Counterfactuals and policy analysis in structural models.
\newblock {\em arXiv preprint arXiv:1302.4929}.

\bibitem[Cheng, 1997]{cheng1997covariation}
Cheng, P.~W. (1997).
\newblock From covariation to causation: A causal power theory.
\newblock {\em Psychological review}, 104(2):367.

\bibitem[Dawid et~al., 2017]{dawid2017}
Dawid, P., Musio, M., and Murtas, R. (2017).
\newblock The probability of causation.
\newblock {\em Law, Probability and Risk}, (16):163--179.

\bibitem[Galles and Pearl, 1998]{galles1998axiomatic}
Galles, D. and Pearl, J. (1998).
\newblock An axiomatic characterization of causal counterfactuals.
\newblock {\em Foundations of Science}, 3(1):151--182.

\bibitem[Glymour, 2013]{glymour2013psychological}
Glymour, C. (2013).
\newblock Psychological and normative theories of causal power and the
  probabilities of causes.
\newblock {\em arXiv preprint arXiv:1301.7377}.

\bibitem[Halpern, 2000]{halpern2000axiomatizing}
Halpern, J.~Y. (2000).
\newblock Axiomatizing causal reasoning.
\newblock {\em Journal of Artificial Intelligence Research}, 12:317--337.

\bibitem[Khoury et~al., 1989]{khoury1989measurement}
Khoury, M.~J., Flanders, W.~D., Greenland, S., and Adams, M.~J. (1989).
\newblock On the measurement of susceptibility in epidemiologic studies.
\newblock {\em American Journal of Epidemiology}, 129(1):183--190.

\bibitem[Koller and Friedman, 2009]{koller2009probabilistic}
Koller, D. and Friedman, N. (2009).
\newblock {\em Probabilistic Graphical Models: Principles and Techniques}.
\newblock MIT press.

\bibitem[Li and Pearl, 2019]{li2019unit}
Li, A. and Pearl, J. (2019).
\newblock Unit selection based on counterfactual logic.
\newblock In {\em Proceedings of the 28th International Joint Conference on
  Artificial Intelligence}, pages 1793--1799. AAAI Press.

\bibitem[Mueller and Pearl, 2020]{mueller2020}
Mueller, S. and Pearl, J. (2020).
\newblock {Which Patients are in Greater Need}: A counterfactual analysis with
  reflections on {COVID-19}.
\newblock \url{https://ucla.in/39Ey8sU+}.

\bibitem[Pearl, 1993]{pearl1993aspects}
Pearl, J. (1993).
\newblock Aspects of graphical models connected with causality.
\newblock {\em Proceedings of the 49th Session of the International Statistical
  Institute, Italy}, pages 399--401.

\bibitem[Pearl, 1995]{pearl1995causal}
Pearl, J. (1995).
\newblock Causal diagrams for empirical research.
\newblock {\em Biometrika}, 82(4):669--688.

\bibitem[Pearl, 1999]{pearl1999probabilities}
Pearl, J. (1999).
\newblock {Probabilities of Causation: Three counterfactual interpretations and
  their identification}.
\newblock {\em Synthese}, 121(1-2):93--149.

\bibitem[Pearl, 2009]{pearl2009causality}
Pearl, J. (2009).
\newblock {\em Causality}.
\newblock Cambridge University Press, {Second} edition.

\bibitem[Spirtes et~al., 2000]{spirtes2000causation}
Spirtes, P., Glymour, C.~N., Scheines, R., and Heckerman, D. (2000).
\newblock {\em {Causation, Prediction, and Search}}.
\newblock MIT press.

\bibitem[Tian and Pearl, 2000]{tian2000probabilities}
Tian, J. and Pearl, J. (2000).
\newblock Probabilities of causation: Bounds and identification.
\newblock {\em Annals of Mathematics and Artificial Intelligence},
  28(1-4):287--313.

\end{thebibliography}

\clearpage
\appendix
\section{Proof of Theorem \ref{thm1}}
\begin{proof}
\begin{eqnarray}
\text{PNS} & = & P(y_x,y'_{x'})\nonumber \\
&=& \Sigma_z P(y_x,y'_{x'}|z)\times P(z)
\label{thm5p1}
\end{eqnarray}
From \cite{li2019unit}, we have the $z$-specific PNS as follows:\\
\begin{flushleft}
\begin{eqnarray}
\max\left\{
\begin{array}{c}
0,\\
P(y_x|z)-P(y_{x'}|z),\\
P(y|z)-P(y_{x'}|z),\\
P(y_x|z)-P(y|z)
\end{array}
\right\}\le z\text{-PNS}
\label{inequ111}
\end{eqnarray}
\end{flushleft}
\begin{flushleft}
\begin{eqnarray}
\min\left\{
\begin{array}{c}
P(y_x|z),\\
P(y'_{x'}|z),\\
P(y,x|z)+P(y',x'|z),\\
P(y_x|z)-P(y_{x'}|z)+\\
+P(y,x'|z)+P(y',x|z)
\end{array}
\right\}\ge z\text{-PNS}
\label{inequ222}
\end{eqnarray}
\end{flushleft}
Substituting \ref{inequ111} and \ref{inequ222} into \ref{thm5p1}, theorem \ref{thm1} holds.\\
Note that since we have,\\
\begin{eqnarray}
&&\sum_z \max\{0,P(y_x|z)-P(y_{x'}|z),\nonumber\\
&&P(y|z)-P(y_{x'}|z),P(y_x|z)-P(y|z)\}\times P(z) \nonumber\\
&\ge& \sum_z 0\times P(z) \nonumber \\
&=& 0, \nonumber \\
&&\sum_z \max\{0,P(y_x|z)-P(y_{x'}|z),\nonumber \\
&&P(y|z)-P(y_{x'}|z),P(y_x|z)-P(y|z)\}\times P(z) \nonumber\\
&\ge& \sum_z [P(y_x|z)-P(y_{x'}|z)]\times P(z) \nonumber \\
&=& P(y_x)-P(y_{x'}), \nonumber \\
&&\sum_z \max\{0,P(y_x|z)-P(y_{x'}|z),\nonumber \\
&&P(y|z)-P(y_{x'}|z),P(y_x|z)-P(y|z)\}\times P(z) \nonumber\\
&\ge& \sum_z [P(y|z)-P(y_{x'}|z)]\times P(z) \nonumber \\
&=& P(y)-P(y_{x'}), \nonumber \\
&&\sum_z \max\{0,P(y_x|z)-P(y_{x'}|z),\nonumber \\
&&P(y|z)-P(y_{x'}|z),P(y_x|z)-P(y|z)\}\times P(z)\nonumber\\
&\ge& \sum_z [P(y_x|z)-P(y|z)]\times P(z)\nonumber\\
&=& P(y_x)-P(y),\nonumber
\end{eqnarray}
then the lower bound in theorem \ref{thm1} is guaranteed to be no worse than the Tian-Pearl lower bound in equation \ref{pnslb}. Similarly, the upper bound in theorem \ref{thm1} is guaranteed to be no worse than the Tian-Pearl upper bound in equation \ref{pnsub}. Also note that, since $Z$ does not contain a descendant of $X$, the term $P(y_x|z)$ refers to experimental data under population $z$. 

\end{proof}

\section{Proof of Theorem  \ref{thm2}}

\begin{proof}
Since $Z$ satisfies the back-door criterion, then equations \ref{inequ11} and \ref{inequ22} still hold and $P(y_x|z)=P(y|x,z)$, $P(y_{x'}|z)=P(y|x',z)$, and $P(y'_{x'}|z)=P(y'|x',z)$. We further have,\\
\begin{eqnarray}
&& P(y_x|z)-P(y_{x'}|z) \nonumber \\
& = & P(y|x,z)-P(y|x',z) \nonumber \\
& \ge & [P(y|x,z)-P(y|x',z)] \times P(x|z) \nonumber \\
& = & P(y|x,z) \times P(x|z) - P(y|x',z)\times (1-P(x'|z)) \nonumber \\
& = & P(y,x|z) + P(y,x'|z) - P(y|x',z) \nonumber \\
& = & P(y|z) - P(y|x',z) \nonumber\\
& = & P(y|z) - P(y_{x'}|z)
\label{thm6p1}
\end{eqnarray}
and
\begin{eqnarray}
&&P(y_x|z)-P(y_{x'}|z) \nonumber\\
& = & P(y|x,z)-P(y|x',z) \nonumber \\
& \ge & [P(y|x,z)-P(y|x',z)] \times P(x'|z), \nonumber \\
& = & P(y|x,z) \times (1- P(x|z)) - P(y|x',z)\times P(x'|z) \nonumber \\
& = & P(y|x,z) - P(y,x|z) - P(y,x'|z) \nonumber \\
& = & P(y|x,z) - P(y|z)\nonumber\\
& = & P(y_x|z) - P(y|z).
\label{thm6p2}
\end{eqnarray}

With equations \ref{thm6p1} and \ref{thm6p2}, equation \ref{inequ11} reduces to equation \ref{lb_confounder} in theorem \ref{thm2}.

We also have,\\
\begin{eqnarray}
&& \min \{ P(y_x|z),P(y'_{x'}|z)\} \nonumber \\
& = & \min \{ P(y|x,z),P(y'|x',z)\} \nonumber \\
& \le & P(y|x,z)\times P(x|z) + P(y'|x',z) \times (1-P(x|z)) \nonumber \\
& = & P(y|x,z)\times P(x|z) + P(y'|x',z) \times P(x'|z) \nonumber \\
& = & P(y,x|z) + P(y',x'|z)
\label{thm6p3}
\end{eqnarray}
and
\begin{flalign}
&\min \{ P(y_x|z),P(y'_{x'}|z)\} \nonumber \\
=\ &\min \{ P(y|x,z),P(y'|x',z)\} \nonumber \\
\le\ &P(y|x,z)\times (1-P(x|z)) + P(y'|x',z) \times P(x|z) \nonumber \\
=\ &P(y|x,z)\times (1-P(x|z)) + P(y'|x',z) \times (1-P(x'|z)) \nonumber \\
=\ &P(y|x,z) - P(y,x|z) + P(y'|x',z) - P(y',x'|z) \nonumber \\
=\ &P(y|x,z) - P(y|x',z)   + P(y,x'|z) + P(y',x|z)\nonumber \\
=\ &P(y_x|z) - P(y_{x'}|z)   + P(y,x'|z) + P(y',x|z).
\label{thm6p4}
\end{flalign}
With equations \ref{thm6p3} and \ref{thm6p4}, equation \ref{inequ22} reduces to equation \ref{ub_confounder} in theorem \ref{thm2}.
\end{proof}

\section{Proof of Theorem  \ref{thm3}} 
\begin{proof}
\begin{eqnarray}
&&\text{PNS} \nonumber\\
&=& P(y_x,y'_{x'})\nonumber\\
&=& \Sigma_z \Sigma_{z'} P(y_x,y'_{x'},z_x,z'_{x'}) \nonumber\\
&=& \Sigma_z \Sigma_{z'} P(y_x,y'_{x'}|z_x,z'_{x'})\times P(z_x,z'_{x'}) \nonumber\\
&\le& \Sigma_z \Sigma_{z'} \min\{P(y_x|z_x,z'_{x'}), P(y'_{x'}|z_x,z'_{x'})\}\times\nonumber\\
&&\ \ \ \ \ \ \ \ \ \ \ \ \min\{P(z_x),P(z'_{x'})\}\nonumber\\
&=& \Sigma_z \Sigma_{z'} \min\{P(y_x|z_x), P(y'_{x'}|z'_{x'})\}\times\label{part_mediator_zx}\\
&&\ \ \ \ \ \ \ \ \ \ \ \ \min\{P(z_x),P(z'_{x'})\} \nonumber\\
&=& \Sigma_z \Sigma_{z'} \min\{P(y|z_x,x), P(y'|z'_{x'},x')\}\times\nonumber\\
&&\ \ \ \ \ \ \ \ \ \ \ \ \min\{P(z_x),P(z'_{x'})\} \label{part_mediator_x}\\
&=& \Sigma_z \Sigma_{z'} \min\{P(y|z,x), P(y'|z',x')\}\times\nonumber\\
&&\ \ \ \ \ \ \ \ \ \ \ \ \min\{P(z_x),P(z'_{x'})\}.\nonumber
\end{eqnarray}
Combined with the Tian-Pearl bounds in equations \ref{pnslb} and \ref{pnsub}, theorem \ref{thm3} holds. Note that equation \ref{part_mediator_zx} is due to $Y_x \independent Z_{x'}\ |\ Z_x$ and $Y_{x'} \independent Z_x\ |\ Z_{x'}$. Equation \ref{part_mediator_x} is due to $\forall x, Y_x \independent\ X\ | Z_x$.
\end{proof}
\section{Proof of Theorem  \ref{thm4}}
\begin{proof}
First we show that in graph $G$, if an individual is a complier from $X$ to $Y$, then $Z_x$ and $Z_{x'}$ must have the different values. This is because the structural equations for $Y$ and $Z$ are $f_y(z,u_y)$ and $f_z(x,u_z)$, respectively. If an individual has the same $Z_x$ and $Z_{x'}$ value, then $f_z(x,u_z)=f_z(x',u_z)$. This means $f_y(f_z(x,u_z),u_y)=f_y(f_z(x',u_z),u_y)$, i.e., $Y_x$ and $Y_{x'}$ must have the same value. Thus this individual is not a complier. Therefore,
\begin{eqnarray}
&&\text{PNS} \nonumber\\
&=&P(y_x,y'_{x'}) \nonumber \\
&=&\Sigma_z \Sigma_{z'\ne z}P(y_z,y'_{z'})\times P(z_x,z'_{x'}) \nonumber \\
&\le&\Sigma_z \Sigma_{z'\ne z}\min\{P(y_z),P(y'_{z'})\}\times\nonumber\\
&& \min\{P(z_x),P(z'_{x'})\} \nonumber \\
&=&\Sigma_z \Sigma_{z'\ne z}\min\{P(y|z),P(y'|z')\}\times\nonumber\\
&& \min\{P(z|x),P(z'|x')\}\nonumber
\end{eqnarray}
Combined with the Tian-Pearl bounds in equations \ref{pnslb} and \ref{pnsub}, theorem \ref{thm4} holds.
\end{proof}
\section{Simulation Algorithm}
We used the following algorithm to generate samples and conduct the simulations in section \ref{simulation}:
\begin{algorithm}[H]
    \SetAlgoLined
    \SetKwData{ProbZ}{prob\_z}
    \SetKwData{ProbYXZ}{prob\_y\_xz}
    \SetKwData{LB}{lb}
    \SetKwData{LBGraph}{lb\_graph}
    \SetKwData{UBGraph}{ub\_graph}
    \SetKwData{UB}{ub}
    \SetKwData{CPT}{cpt}
    \SetKwFunction{UnifRand}{random-uniform}
    \SetKwFunction{PNS}{pns-bounds}
    \SetKwFunction{PNSGraph}{pns-graph}
    \SetKwFunction{AppendToResult}{append-result}
    \SetKwFunction{Alg}{generate-cpt}
    \SetKwInOut{Input}{input}\SetKwInOut{Output}{output}
    \Input{Number of output samples $n$\\Causal diagram $G$\\Covariates to condition on $Z$}
    \Output{List of 4-tuples consisting of general lower bound, lower bound with causal graph, upper bound with causal graph, and general upper bound}
    \Begin{
    \For{$i \leftarrow 1$ \KwTo $n$}{
        
        \CPT $\leftarrow$ \Alg{$G$,\UnifRand} \;
    
        \tcp{Lower/upper Tian-Pearl bounds}
        \LB, \UB $\leftarrow$ \PNS{\CPT}\;
        
        \tcp{Lower/upper bounds with graph}
        \LBGraph, \UBGraph $\leftarrow$ \PNSGraph{\CPT, $Z$}\;
        
        \AppendToResult{\LB, \LBGraph, \UBGraph, \UB}\;
    }
    }
    \caption{Generate PNS simulation data with theorem \ref{thm2}}
\end{algorithm}

\begin{procedure}
    \SetAlgoLined
    \SetKwData{ProbZ}{prob\_z}
    \SetKwData{ProbYXZ}{prob\_y\_xz}
    \SetKwData{LB}{lb}
    \SetKwData{LBGraph}{lb\_graph}
    \SetKwData{UBGraph}{ub\_graph}
    \SetKwData{UB}{ub}
    \SetKwData{S}{s}
    \SetKwData{P}{p}
    \SetKwData{Sum}{sum}
    \SetKwFunction{UnifRand}{uniform-random}
    \SetKwFunction{PNS}{pns-bounds}
    \SetKwFunction{PNSGraph}{pns-graph}
    \SetKwFunction{AppendToResult}{append-result}
    \SetKwFunction{Sample}{sample}
    \SetKwFunction{NumberofIns}{num-instantiates}
    \SetKwInOut{Input}{input}\SetKwInOut{Output}{output}
    \Input{$n$ causal diagram nodes ($X_1,...,X_n$)\\Distribution $D$}
    \Output{$n$ conditional probability tables for $P(X_i|Parents(X_i))$}
    \Begin{
    \For{$i \leftarrow 1$ \KwTo $n$}{
        \S $\leftarrow$ \NumberofIns{$X_i$}
        
        \P $\leftarrow$ \NumberofIns{$Parents(X_i)$}
        
        \For{$k \leftarrow 1$ \KwTo $p$}{
            \Sum $\leftarrow$ $0$
            
                \For{$j \leftarrow 1$ \KwTo $s$}{
                    $a_j$ $\leftarrow$ \Sample{$D$}
                    
                    \Sum $\leftarrow$ \Sum $+$ $a_j$
                }
            
                \For{$j \leftarrow 1$ \KwTo $s$}{
                    $P(x_{i_j}|Parents(X_i)_k)$ $\leftarrow$ $a_j/$\Sum

                }
        }
    }
    }
    \caption{generate-cpt()}
\end{procedure}
For figure \ref{causalgexample3}, binary variables $Z_1$ and $Z_2$ were considered as a covariate with $4$ instantiates. Similarly, figure \ref{causalgexample4}'s $Z$ variables were considered as a single covariate with $8$ instantiates.

\end{document}